\documentclass[a4paper,11pt]{article}
\usepackage{jcappub} 
\usepackage{ulem}
\usepackage{graphicx}
\usepackage{graphicx}
\usepackage{epstopdf}
\usepackage{dcolumn}
\usepackage{bm}
\usepackage{comment}
\usepackage[utf8]{inputenc}
\usepackage{mciteplus}
\usepackage{color}
\usepackage{subfigure}
\usepackage{comment}
\usepackage{enumerate}
\usepackage{mathrsfs}
\usepackage{units}
\usepackage{hyperref,graphicx,amsfonts,amssymb,amsthm,amsmath,psfrag}
\usepackage[toc,page]{appendix}
\usepackage{subfigure}
\usepackage{slashed}
\usepackage{tikz}
\usepackage{booktabs}
\usepackage{siunitx}
\usepackage{multirow}
\usepackage{color}
\usepackage{todonotes}

\usetikzlibrary{decorations.markings,snakes}
\tikzset{
  fermionline/.style={line width=1pt,postaction={decorate},
    decoration={markings,
      mark=at position 0.5 with {\draw[-stealth] (0,0)--(2pt,0);}}},
  bosonline/.style={line width=1pt,decorate,
    decoration={snake,amplitude=1,segment length=4}},
  higgsline/.style={line width=1pt,dashed}
}
\newcommand{\be}{\begin{equation} }
\newcommand{\ee}{\end{equation}}

\newcommand{\lmax}{\mathrm{max}}
\newcommand{\lmin}{\mathrm{min}}
\renewcommand{\arraystretch}{1.2}
\usepackage{amsmath}

\title{
\huge{Preheating in Palatini Higgs inflation}}
\author{Javier Rubio and}
\author{Eemeli S. Tomberg}
\emailAdd{javier.rubio@helsinki.fi} 
\emailAdd{eemeli.tomberg@helsinki.fi}
\vspace{5cm}
\affiliation{Department of Physics and Helsinki Institute of Physics, \\  PL 64, FI-00014 University of Helsinki, Finland} 

\abstract{We study the details of preheating in Palatini Higgs inflation. We show, that contrary to what happens in the metric formulation of the model, the Universe does not reheat through the creation of gauge bosons only, but also through the tachyonic production of Higgs excitations. The latest entropy production channel turns out to be very efficient and leads to an almost instantaneous onset of radiation domination after the end of inflation. As compared to the metric case, this reduces the number of e-folds needed to solve the
usual hot big bang problems while leading to a smaller spectral index for the primordial spectrum of density perturbations.} 
\preprintnumber{HIP-2019-4/TH}
\begin{document}
\maketitle

\section{Introduction}
Higgs inflation is an appealing inflationary model based on the inclusion of a non-minimal coupling between the Standard Model Higgs field and gravity \cite{Bezrukov:2007ep} (for a recent review and extensive list of references see Ref.~\cite{Rubio:2018ogq}). Although the scenario was initially formulated as a metric theory in which the connection determining the Ricci scalar was identified with the  Levi-Civita connection \cite{Bezrukov:2007ep,GarciaBellido:2008ab,Bezrukov:2008ut,Bezrukov:2009db,Burgess:2009ea,Barbon:2009ya,Burgess:2010zq,Hertzberg:2010dc,Atkins:2010yg,Bezrukov:2010jz,Bezrukov:2011sz,Bezrukov:2014bra,Bezrukov:2014ipa,Rubio:2015zia, DeCross:2015uza, Repond:2016sol,DeCross:2016fdz, DeCross:2016cbs,Ema:2016dny,Fumagalli:2016lls,Enckell:2016xse,Bezrukov:2017dyv,Sfakianakis:2018lzf}, this need not be the case. In particular, one could consider an alternative Palatini formulation of gravity in which the metric and the connection are taken to be independent geometrodynamical variables \cite{Bauer:2008zj,Bauer:2010jg,Rasanen:2017ivk,Markkanen:2017tun,Enckell:2018kkc,Rasanen:2018fom,Rasanen:2018ihz,Takahashi:2018brt,Tenkanen:2019jiq}. As compared to the metric case, this formulation displays some interesting features. First, it leads to different inflationary predictions \cite{Bauer:2008zj}, opening the door to test the nature of gravity with future cosmological observations. Second, it raises the effective cutoff of the theory without introducing additional degrees of freedom below the Planck scale \cite{Bauer:2010jg}. Third, it does not require the introduction of the usual Gibbons–Hawking–York term in order to  obtain  the  equations  of  motion \cite{Ferraris1982}. Finally, it leads to different interactions among the Higgs field and the Standard Model particles in the large field regime, with a potential impact on the entropy production process following the end of inflation. 

In this paper we study the preheating stage in Palatini Higgs inflation, highlighting the differences with the well--studied metric case \cite{GarciaBellido:2008ab,Bezrukov:2008ut,Repond:2016sol,DeCross:2015uza, DeCross:2016fdz, DeCross:2016cbs, Ema:2016dny, Sfakianakis:2018lzf}.\footnote{A Jordan-frame analysis of preheating in Palatini theories involving non-minimally couplings to gravity was performed in Ref.~\cite{Fu:2017iqg}. The analysis presented there was, however, completely unrelated to the Higgs field and limited to non-minimal couplings to gravity much smaller than those considered in this paper.} We show that the main mechanisms leading to the onset of the hot big bang in these two alternative scenarios are indeed very different. While the depletion of the inflaton condensate in the metric case is driven by the production of electroweak gauge bosons via \textit{parametric resonance} \cite{Dolgov:1989us,Traschen:1990sw,Kofman:1994rk,Shtanov:1994ce,Kofman:1997yn,Greene:1997fu}, the Palatini one is dominated by the production of Higgs excitations via \textit{tachyonic preheating} \cite{Felder:2000hj,Felder:2001kt}. As compared to the metric case, the entropy production in Palatini Higgs inflation turns out to be significantly more efficient, reducing the number of inflationary e-folds needed to solve the flatness and horizon problems and leading to a smaller spectral tilt for primordial density perturbations.

This paper is organized as follows. The inflationary and post-inflationary dynamic of the metric and Palatini formulations of Higgs inflation is presented in Section \ref{sec:metricvsPalatini}. After reviewing  in Section \ref{sec:metricpre} the preheating stage in the metric scenario, we present in Section \ref{sec:particle_production} a detailed analysis of the entropy production in Palatini Higgs inflation. The impact of this stage on cosmological observables is discussed in Section \ref{sec:impact}. Section \ref{sec:conclusions} contains our conclusions. 

\section{Metric vs Palatini}\label{sec:metricvsPalatini}

As formulated in the Standard Model, the Higgs field is not a suitable inflaton candidate. In particular, its self-interaction significantly exceeds the value needed to solve the hot big bang problems without generating an excessively large amount of primordial density perturbations.  In Higgs inflation, this difficulty is overcome by introducing a non-minimal coupling of the Higgs field to gravity. When written in the unitary gauge the graviscalar part of the Higgs inflation action takes the form  \cite{Bezrukov:2007ep}
\begin{equation} \label{eq:nonminimal_action}
S = \int d^4x \sqrt{-g} \left[\frac{M_{P}^2 + \xi h^2}{2} g^{\mu\nu}R_{\mu\nu}(\Gamma) - \frac{1}{2} g^{\mu\nu}\partial_{\mu}h\partial_{\nu}h - V(h) \right] \,,
\end{equation}
with  
\begin{equation}\label{eq:connect}
 R_{\mu\nu}= \partial_\sigma  \Gamma_{\ \mu\nu}^\sigma
-\partial_\mu \Gamma^\sigma_{\ \sigma\nu}
+  \Gamma^\rho_{\ \mu\nu} \Gamma^\sigma_{\ \sigma\rho}- \Gamma^\rho_{\ \sigma\nu}\Gamma^\sigma_{\ \mu\rho}\,
\end{equation}
the Ricci tensor and
\begin{equation}
V(h)=\frac{\lambda}{4}(h^2-v^2)^2 
\end{equation}
the usual Standard Model Higgs potential. The connection
 $\Gamma^\rho{}_{\mu\nu}$ in Eqs.~\eqref{eq:nonminimal_action} and \eqref{eq:connect} is taken to be the Levi-Civita connection 
   \begin{equation}
{\Gamma}^{\sigma}{}_{\mu\nu} = \frac{1}{2} g^{\sigma \rho}\left( \partial_{\mu} g_{\nu \rho} + \partial_{\nu} g_{\rho \mu} - \partial_{\rho} g_{\mu\nu} \right)
        \end{equation}
 in the metric formulation of gravity and an arbitrary one in its Palatini  counterpart, being the only assumption that it is torsion-free, i.e. $\Gamma^\rho{}_{\mu\nu}=\Gamma^\rho{}_{\nu\mu}$. 
 
 The differences among metric and Palatini formulations are more easily understood in the so-called Einstein frame. This is achieved by performing a Weyl rescaling of the metric, $g_{\mu\nu} \rightarrow  \Omega^{-2}(h) g_{\mu\nu}$, with  Weyl factor $\Omega^2(h)=1+\xi h^2/M_{P}^2$. 
While  this transformation modifies the Ricci scalar in the metric formulation, it leaves it invariant in the Palatini case, where the metric and the connection are completely unrelated. This translates into an Einstein-frame action displaying a different kinetic structure for the metric ($\alpha= 1+6\xi$) and Palatini ($\alpha=0$) cases, namely
\begin{equation}
\label{higgsLagrangianE2}
 S_E= \int d^4x \sqrt{-{g}} \, \left[   \frac{M_P^2}{2}R
    -\frac{1}{2}\left(\frac{1+\alpha \, \xi^2 h^2/M_P^2}{(1+\xi h^2/M_P^2)^2}\right)
     g^{\mu\nu}\partial_\mu h\, \partial_\nu h - U(h)\right]\, ,
\end{equation}
with $U(h)\equiv V(h)\Omega^{-4}(h)$ a Weyl-rescaled Higgs potential. The non-canonical kinetic term for the $h$ field in this expression can be made canonical by performing a field redefinition
\begin{equation} \label{relchih}
  \frac{d\chi}{dh}=
  \sqrt{\frac{1+\alpha \, \xi h^2/M_P^2 }{ (1+\xi h^2/M_P^2)^2}}\,,
  \end{equation}
which can be easily integrated to obtain \cite{GarciaBellido:2008ab}
\begin{equation}   \label{relchihInt}
\frac{\sqrt{\xi} \chi}{M_P} = \sqrt{\alpha} \,\mathrm{arsinh}\left(\frac{\sqrt{\alpha\xi}h}{M_P}\right) - \sqrt{\alpha-1} \,\mathrm{artanh}\left( \frac{\sqrt{(\alpha-1)\xi} h/M_P}{\sqrt{1+\,\alpha\,\xi h^2/M_P^2}} \right) \,.
\end{equation}
When written in terms of $\chi$, the Einstein-frame action \eqref{higgsLagrangianE2} takes the form 
\begin{equation} \label{eq:nonminimal_action_einstein}
	S_E = \int d^4x \sqrt{-g} \left[ \frac{M_{P}^2}{2} g_E^{\mu\nu}R_{\mu\nu}(\Gamma) - \frac{1}{2} g_E^{\mu\nu}\partial_{\mu}\chi\partial_{\nu}\chi - U(\chi) \right] \,.
\end{equation}
The difference between metric and Palatini formulations is now encoded in the Einstein-frame potential
\begin{equation}\label{eq:Upot}
U(\chi) \equiv \frac{V(h(\chi))}{\Omega^{4}(h(\chi))}\equiv  \frac{\lambda}{4}F^4(\chi)\,,
\end{equation}
where
  \begin{equation}\label{Ftree}
 F(\chi)\simeq
\begin{cases} 
\chi & \textrm{for} \hspace{1mm} \chi< \frac{M_P}{\xi}\,, \\
\frac{M_P}{\sqrt{\xi}} \left(1-e^{-\sqrt{\frac{2}{3}}\frac{\chi}{M_P}}\right)^{\frac12} 
 &   \textrm{for} \hspace{1mm}\chi> \frac{M_P}{\xi}\,,
   \end{cases}
\end{equation}
for the metric case and\footnote{This expression is \textit{exact}.} 
\begin{equation}\label{FtreeP}
	F(\chi) = \frac{\lambda M_P}{\sqrt{\xi}}\tanh \left( \frac{\sqrt{\xi}\chi}{M_P} \right)\,,
 \end{equation}
for the Palatini one. At large field values $\chi \gg M_P/\sqrt{\xi}$ both \eqref{Ftree} and \eqref{FtreeP} become approximately constant, leading to an exponentially flat potential suitable for inflation with the usual slow-roll initial conditions. Using the standard techniques, we get the following predictions for the tilt of the primordial spectrum of curvature perturbations $n_s$, its amplitude $A_s$ and the tensor-to-scalar ratio $r$ \cite{Bezrukov:2007ep,Bauer:2008zj},
\begin{equation} \label{eq:cmb_observables}
n_s \simeq 1 - \frac{2}{N_*} \, , \hspace{15mm}
A_s \simeq
 \begin{cases} 
\frac{\lambda N_*^2}{72 \pi^2\xi^2} &  \hspace{4mm} \textrm{metric}\,, \\
\frac{\lambda N_*^2}{12\pi^2\xi } & \hspace{4mm} \textrm{Palatini}\,,
   \end{cases} 
\hspace{15mm} r \simeq
 \begin{cases} 
\frac{12}{N_*^2}  &  \hspace{4mm} \textrm{metric}\,, \\
\frac{2}{\xi N_*^2} & \hspace{4mm} \textrm{Palatini}\,,
   \end{cases} 
\end{equation}
with $N_*$  the number of e-folds at which the pivot scale $k_* = 0.05$ Mpc${}^{-1}$ crosses the horizon. Note that, although the functional dependence of the spectral tilt is the same in the two formulations, the amplitude $A_s$ and the tensor--to--scalar ratio $r$ are different. This has some interesting consequences. In particular, when compared to the observed value  $A_S \approx 2.1 \times 10^{-9}$ \cite{Akrami:2018odb},  the two spectral amplitudes translate into different relations among the non-minimal coupling $\xi$ and the Higgs self-coupling $\lambda$, namely
\begin{equation} \label{eq:cmb_derived_quantities}
\xi \simeq  \begin{cases} 
 800 N_*\sqrt{\lambda} &  \hspace{5mm} \textrm{metric}\,, \\
3.8 \times 10^{6} N_*^2\lambda 
 & \hspace{5mm} \textrm{Palatini}\,.
   \end{cases} 
\end{equation}
Assuming that the renormalization group running of the Standard Model parameters from the electroweak scale does not give rise to excessively fine-tuned values of $\lambda$ at the inflationary scale, the above expressions require the non-minimal coupling $\xi$ to be large in both metric and Palatini formulations. For inflationary values $\lambda=10^{-4}\dots0.1$ and $N_*\approx 50$, we get $\xi=100\dots10^5$ in the metric case and $\xi=10^6\dots10^9$ in the Palatini one.
The value of the tensor--to--scalar ratio following from the combination of these numbers with Eq.~\eqref{eq:cmb_observables} is significantly smaller in the Palatini formulation, allowing to exclude the model in the eventual detection of a tensor--to--scalar ratio within the accuracy of future cosmological observations \cite{Martin:2014rqa}. 

Metric and Palatini formulations display also different post-inflationary evolutions. To see this explicitly, let us consider the equations of motion for the inflaton field in a Friedmann-Lema\^itre-Robertson-Walker background, namely 
\begin{equation}
\label{eq:background_eom}
	\ddot{\chi} + 3H\chi + U'(\chi) = 0 \, , 
	\hspace{10mm}
	H^2 =\frac{1}{3M_P^2}\left( \frac{1}{2}\dot{\chi}^2 + U(\chi)\right) \, , \hspace{10mm} \quad \dot{H} = -\frac{\dot{\chi}^2}{2 M_P^2} \, ,
\end{equation}
with  $H=\dot a/a$ the Hubble rate, $a$ the scale factor and the dots denoting derivatives with respect to the cosmic time $t$. Combining these equations, we can write
\begin{equation}\label{Hrate}
	M_P^2 \frac{dH}{dh} = -\frac{\dot{\chi}^2}{2\dot{h}} = -\frac{1}{2}\sqrt{6H^2 M_P^2 - 2U[\chi(h)]}\frac{d\chi}{dh} \, ,
\end{equation}
with $U$ having the same $h$-dependence in both metric and Palatini formulations. A simple inspection of Eq.~\eqref{relchih} reveals  that $d\chi/dh$ is always bigger in the metric case. In particular, for intermediate field values $M_P/\xi \ll h \ll M_P/\sqrt{\xi}$ we have $d\chi/dh\approx 1$ in the Palatini case, but $d\chi/dh\approx \sqrt{6}\xi h/M_P \gg 1$ in the metric one. When combined with Eq.~\eqref{Hrate} this implies that after the end of inflation (happening at $h \sim M_P/\sqrt{\xi}$ in both formulations) the Hubble rate in the metric case decreases much faster as $h$ approaches zero. Since this quantity determines the total energy of the system, the oscillations of the $\chi$-field are damped less effectively in the Palatini scenario. As we will show below, this has important consequences.

\section{Preheating in metric Higgs inflation}\label{sec:metricpre}

The preheating stage in metric Higgs inflation has been extensively studied in the literature \cite{GarciaBellido:2008ab,Bezrukov:2008ut,Repond:2016sol,DeCross:2015uza, DeCross:2016fdz, DeCross:2016cbs, Ema:2016dny, Sfakianakis:2018lzf}. The rapid decrease of the Hubble rate soon after the end of inflation reduces the amplitude of the inflaton field to a range $M_P/\xi<\chi <\sqrt{3/2} M_P$ where the effective potential following from Eq.~\eqref{Ftree} can be safely approximated by a polynomial form \cite{GarciaBellido:2008ab,Bezrukov:2008ut}
\begin{equation}\label{metricquadraticP} 
U(\chi)\simeq \frac{1}{2}{\cal M}^2\chi^2\,,\hspace{10mm} \textrm{with} \hspace{10mm}  {\cal M} =\sqrt{\frac{\lambda}{3}} \frac{M_P}{\xi}\,.
\end{equation} 
The absence of Higgs self-interactions in this regime together with the Pauli blocking of fermions \cite{Greene:1998nh,Giudice:1999fb,GarciaBellido:2000dc,Greene:2000ew,Peloso:2000hy,Berges:2010zv} suppresses the direct production of the associated excitations. The main mechanism draining the energy of the Higgs condensate is the non-perturbative production of gauge bosons at the bottom of the potential. Upon creation, these particles tend to decay into the Standard Model fermions with a large decay rate proportional to their field-dependent mass, restraining with it the onset of parametric resonance \cite{GarciaBellido:2008ab,Bezrukov:2008ut,Repond:2016sol}, see also Refs.~\cite{Kasuya:1996np,Mukaida:2012bz}. Since the amplitude of the Higgs field decreases with time due to the expansion of the Universe, this blocking effect is only temporarily active. Eventually the decay rate into fermions becomes small enough as to allow the gauge bosons to accumulate, leading to a strong depletion of the Higgs condensate \cite{GarciaBellido:2008ab,Bezrukov:2008ut,Repond:2016sol}. When the energy density into Standard Model particles becomes comparable to the background component, the resonant production of gauge bosons terminates due to backreaction effects. From there on, the energy transfer from the inflaton field to the Standard Model particles continues through a slower turbulent stage where the total energy of the Universe becomes democratically distributed among the different species \cite{Micha:2002ey,Micha:2004bv,Repond:2016sol}. 

Note that the above treatment does not account for the differences between transverse and longitudinal degrees of freedom. As noticed in Refs.~\cite{DeCross:2015uza, DeCross:2016fdz, DeCross:2016cbs, Ema:2016dny, Sfakianakis:2018lzf}, the longitudinal gauge boson components could be explosively produced by a `Riemannian spike' appearing in the Higgs-doublet Einstein-frame kinetic manifold at small field values. In practice, the longitudinal gauge bosons acquire mass terms proportional to the time derivative of the field $h$, which can be written as
\begin{equation} \label{eq:bg_time_derivs}
	\dot{h} = \frac{dh}{d\chi}\dot{\chi} \,,
\end{equation}
with $dh/d\chi$ given by  Eq.~\eqref{relchih}. 
Although the Einstein-frame velocity $\dot{\chi} \sim H^2 M_P^2$ behaves smoothly everywhere, the differential field relation $dh/d\chi$ changes rapidly at the transition value $h\simeq M_P/\xi$ and hence does $\dot h$ and the longitudinal gauge boson mass, leading to a strong adibaticity violation and the consequent production of particles.  Although it has been argued that this effect could lead to a full depletion of the Higgs condensate in a single oscillation \cite{Ema:2016dny}, this conclusion should be taken with care. In particular, the mass of the longitudinal gauge bosons blows up only for an extremely short interval of time. By the uncertainty principle, this translates into the creation of highly energetic particles with a momentum scale far beyond the cutoff of the theory at those field values \cite{Bezrukov:2010jz}. In the absence of an ultraviolet completion, it seems difficult to conclude whether the explosive production of longitudinal modes is a true physical effect or just the artifact of an out--of--control quantum treatment. Remarkably enough, even though the spiky behaviour persists in extensions of metric Higgs inflation leading to a higher cutoff scale \cite{Ema:2017rqn,Gorbunov:2018llf}, the preheating stage in those scenarios cannot be completed within a single oscillation \cite{He:2018mgb}.

\section{Preheating in Palatini Higgs inflation} \label{sec:particle_production}

\begin{figure}
\begin{center}
\includegraphics[scale=1]{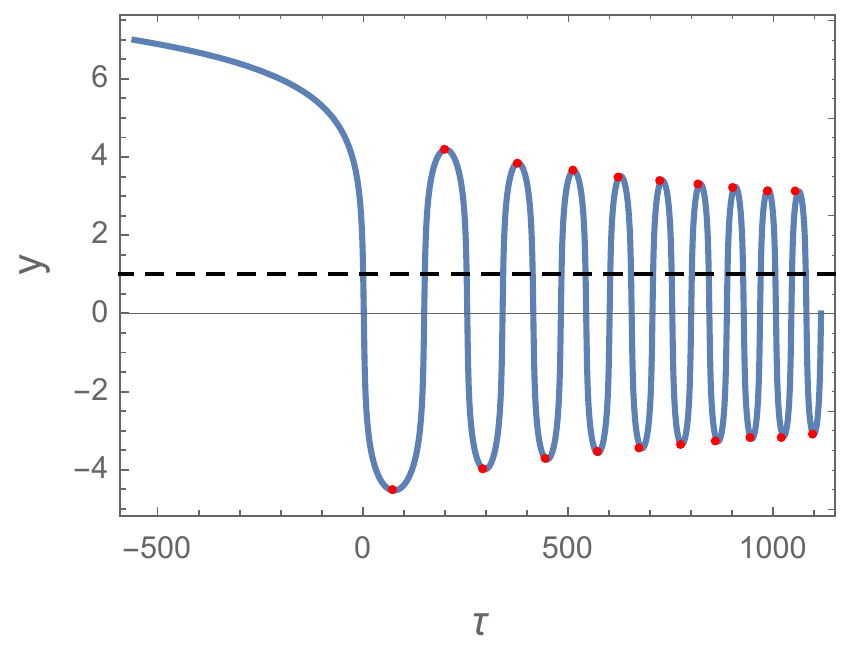}
\end{center}
\caption{Evolution of the inflaton field in Palatini Higgs inflation. The dashed horizontal line at $y=1$ signals the transition to the plateau regime. The red dots indicates the maximal amplitude per oscillation $y_{max}$ following from our analytical estimates. Note that, in the absence of any other damping effect beyond the expansion of the Universe, the oscillating field $y$ is able to return to the plateau, $y_{\rm max}\gtrsim 1$, for a large number of oscillations.}
\label{fig:evolution}
\end{figure}

Although the quartic structure of the Palatini Higgs potential at small field values might lead us to believe that the Universe would expand as a radiation-dominated one during the (p)reheating stage, this is surprisingly not the case due to the mild damping of the background oscillations advocated in Section \ref{sec:metricvsPalatini}. To see this explicitly, let us consider the background evolution equations \eqref{eq:background_eom} in the Palatini case, i.e. with 
\begin{equation} \label{eq:potential_einsteinP}
U(\chi) = \frac{\lambda M_P^4}{4\xi^2}\tanh^4 \left( \frac{\sqrt{\xi}\chi}{M_P}\ \right)\,.
\end{equation}
In order to minimize the dependence on the couplings it is convenient to recast these equations in terms of rescaled quantities
\begin{equation} \label{eq:rescaled_variablesB}
\begin{split}
	y \equiv \frac{\sqrt{\xi}\chi}{M_P} \, , \hspace{20mm}
	E \equiv \frac{\xi}{\sqrt{\lambda} M_P} H \, , \hspace{20mm}
	\tau \equiv \sqrt{\frac{\lambda}{\xi}} M_P \,t \,. 
\end{split}
\end{equation}
We get
\begin{equation}
\label{eq:background_eom_rescaled}
	y'' + 3\frac{E}{\sqrt{\xi}} y' + u'(y) = 0 \, , \hspace{15mm}
	3E^2 = \frac{1}{2}y'{}^2 + u(y) \, ,  \hspace{15mm} E' = -\frac{y'{}^2}{2\sqrt{\xi}} \, , 
\end{equation}
with 
\begin{equation}\label{dimlesspot}
u(y) = \frac{1}{4}\tanh^4 y \,,
\end{equation}
and the primes denoting derivatives with respect to $\tau$. The numerical solution of these equations with initial  slow-roll conditions\footnote{In terms of the dimensionful field variable $\chi$, this corresponds to initial conditions 
\begin{equation} \nonumber
\begin{split}
	\dot{\chi} = -\frac{U'(\chi_i)}{3H} \, , \hspace{15mm}
	H = \sqrt{\frac{U(\chi_i)}{3 M_P^2}} 
	\, ,\hspace{10mm} \textrm{at} \hspace{10mm} \chi = \chi_i\gg \frac{M_P}{\sqrt{\xi}}\,.
\end{split}
\end{equation}}
\begin{equation} \label{eq:inital_conditions_rescaled}
\begin{split}
	y' = -\frac{\sqrt{\xi} u'(\chi_i)}{3E} \, , \hspace{20mm}
	E = \sqrt{\frac{u(y_i)}{3}} \approx \frac{1}{\sqrt{12}} \equiv E_i \, , 
\end{split}
\end{equation}
at $y = y_i \gg 1$ is presented in Fig.~\ref{fig:evolution}. As shown in this plot, the amplitude and energy of the scalar field during the oscillatory stage remains approximately constant for a large number of oscillations. 
For typical amplitudes $y_\lmax>1$, the field $y$ returns periodically to the plateau region of the potential and spends most of its time there, such that the average expectation value of its potential is approximately constant, $\langle u(y) \rangle \approx 3E^2$. The associated equation--of--state  parameter is therefore close to a de Sitter value, 
\begin{equation} \label{eq:eq_of_state}
	w \equiv \frac{\langle p \rangle}{\langle \rho \rangle} = \frac{3E^2-2 \langle u(y) \rangle}{3E^2} \approx -1 \, ,
\end{equation}
meaning that the exponential expansion of the Universe continues also during the oscillatory phase. To get some additional insight on this rather non-intuitive result we will follow a perturbative approach. Neglecting the friction term in Eq.~\eqref{eq:background_eom_rescaled} during an oscillation,\footnote{The friction term is important during slow-roll inflation and becomes smaller afterwards. Note indeed that $\xi$ is large and $E \sim E_i= 1/\sqrt{12}$ at the onset of the oscillations, so the friction term in this phase is small compared to the other terms in Eq.~\eqref{eq:background_eom_rescaled}.

}
we obtain a zero-th order equation of motion
\begin{equation} \label{eq:inflaton_eom_no_friction}
	y'' + u'(y) \approx 0 \,,
\end{equation}
preserving the energy of the system and the initial amplitude of the field. For $y\leq 1$, the field oscillates around the potential minimum $y=0$ with semiperiod
\begin{gather} \label{eq:semiosc_time}
	\Delta \tau = \int_{-y_\lmax}^{y_\lmax} \frac{dy}{y'}
	 = \int_{-y_\lmax}^{y_\lmax} \frac{dy}{\sqrt{6E^2 - 2u(y)}}
	 = \frac{1}{\sqrt{6E^2}} I_T(y_\lmax) \, ,
\end{gather}
where\footnote{In the last step, we have made use of elliptic integral properties and dropped sub-leading terms for big $z$, namely
\begin{equation}\nonumber
\frac{d I_T(z)}{d e^z} \approx \frac{I_T(z)}{e^{z}} \hspace{10mm}\Longrightarrow \hspace{10mm} I_T(z) \approx e^z\lim_{z \to \infty}\frac{ I_T(z)}{e^z} =\frac{ \pi}{2\sqrt{2}} e^z \,.
\end{equation}
}
\begin{gather} 
\label{eq:IT} 
	 I_T(z) \equiv 2\int_{0}^{z} dx \left( 1 - \frac{\tanh^4 x}{\tanh^4 z} \right)^{-\frac12} = 2 \,\Pi(\tanh^2 z, i) \tanh z \simeq \frac{ \pi e^z}{2\sqrt{2}}  
\end{gather}
and $\Pi$ denotes the complete elliptic integral of third kind. The perturbative change of the energy density due to the omitted friction term can be calculated from Eq.~\eqref{eq:background_eom_rescaled}. We get
\begin{gather}
\label{eq:E_change}
	 \Delta E = -\frac{1}{2\sqrt{\xi}} \int d\tau \, y'{}^2 = -\frac{1}{2\sqrt{\xi}} \int_{-y_\lmax}^{y_\lmax} dy \sqrt{6E^2 - 2u(y)} = -\frac{E}{2}\sqrt{\frac{6}{\xi}} I_H(y_\lmax) \, , 
\end{gather}	 
with  
\begin{gather}
\label{eq:IH}
	 I_H(z) \equiv 2\int_{0}^{z} dx \left( 1 - \frac{\tanh^4 x}{\tanh^4 z} \right)^{1/2} \,
\end{gather}
a function that can be either expressed in terms of elliptical integrals or calculated numerically.

We can use a similar technique to calculate the oscillation amplitude after the first zero crossing. The field starts at $y=\infty$ with $E=E_i$, crosses zero and rolls to a value $-y_\lmax$, where it turns around. At $-y_\lmax$, the field is momentarily at rest, so all its energy is in the potential, i.e. $E=\sqrt{u(y_\lmax)/3}$. The energy change between $y=\infty$ and $-y_\lmax$ is then given by
\begin{equation} \label{eq:initial_ymax_condition0}
\begin{split}
\Delta E_i &= \sqrt{\frac{u(y_\lmax)}{3}} - E_i 
=  -\frac{1}{2\sqrt{\xi}} \int_{-y_\lmax}^{\infty} dy \sqrt{6E^2 - 2u(y)} \\
&\approx -\frac{1}{2\sqrt{\xi}} \int_{-\infty}^{\infty} dy \sqrt{6E_i^2 - 2u(y)} = -\frac{E_i}{2}\sqrt{\frac{6}{\xi}} I_H(\infty) \, .
\end{split}
\end{equation}
To solve for $y_\lmax$, let us expand the potential $u(y)$ in terms of a new variable $\delta \equiv 4 e^{-2y}$,\footnote{This is chosen so that $\delta \approx M_P^2/(\xi h^2)$ for $\delta < 1$.} which turns out to be small for all the $y_\lmax$ values we need to consider. To leading order in the quantity, we can approximate
\begin{equation}\label{eq:deltaexp}
\sqrt{\frac{u(y_\lmax)}{3}} - E_i \approx -\delta\, E_i \,.
\end{equation}
Combining this expression with Eq.~\eqref{eq:initial_ymax_condition0} and solving for $\delta$ we get
\begin{gather}
\label{eq:initial_ymax}
	\delta_\lmax \approx \sqrt{\frac{6}{\xi}} \frac{I_H(\infty)}{2} \, , \hspace{20mm}
	y_\lmax \approx -\frac{1}{2} \log \left( \frac{\delta_\lmax}{4} \right) \, , 
\end{gather}
with 
\begin{gather}
\label{eq:IH_infty}
	I_H(\infty) = \frac{\Gamma^2\left(1/4 \right)}{2\sqrt{2\pi}} + 2 \, \mathrm{Im} \, \mathcal{E}(2) \, \approx 3.8202 \, ,
\end{gather}
and $\mathcal{E}$ the complete elliptic integral of the second kind. For non-minimal couplings in the range $\xi = 10^6 \dots 10^9$, we have $\delta_\lmax=1.5 \times 10^{-4} \dots 4.7 \times 10^{-3} \ll 1$, $y_\lmax=3.4 \dots 5.1$ and $(E_i-E)/E_i= 1.5 \times 10^{-4} \dots 4.7\times 10^{-3} \ll 1$ at $y_\lmax$. For the subsequent oscillations, Eq.~\eqref{eq:E_change} gives $\Delta E/E = 1.5 \times 10^{-4} \dots 4.3\times 10^{-3} \ll 1$, showing that the energy of the system changes indeed slowly and that the inflaton field is able to return to the inflationary plateau for hundreds of oscillations.\footnote{This interval becomes longer for increasing $\xi$ values, since they decrease the friction term in Eq.~\eqref{eq:background_eom_rescaled}.} This implies in turn that, contrary to  what happens in the metric case, the potential \eqref{eq:potential_einsteinP} cannot be approximated by a simple polynomial expansion soon after the end of inflation.

\subsection{Higgs production} \label{sec:higgs_production}
\begin{figure}
\begin{center}
\includegraphics[scale=0.8]{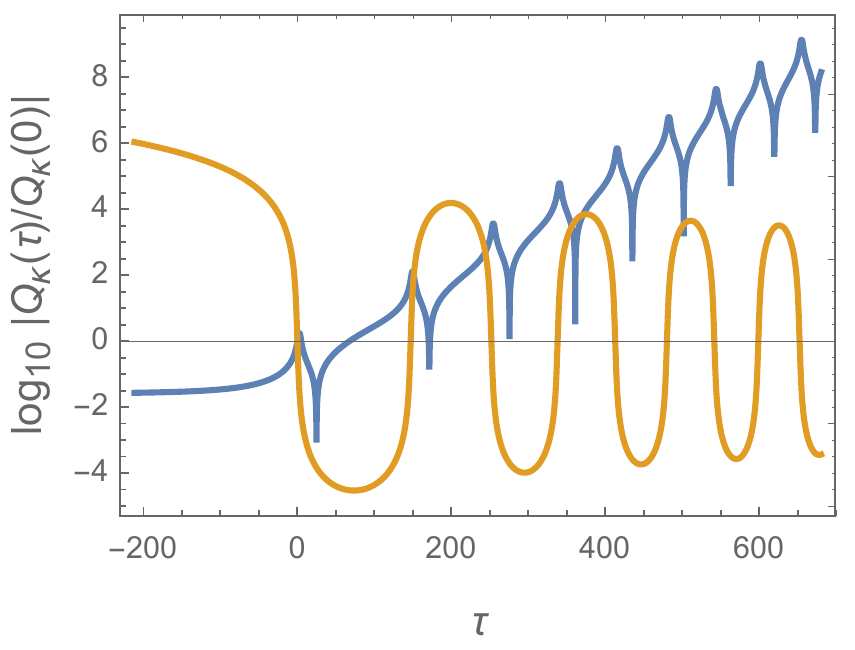}
\includegraphics[scale=0.8]{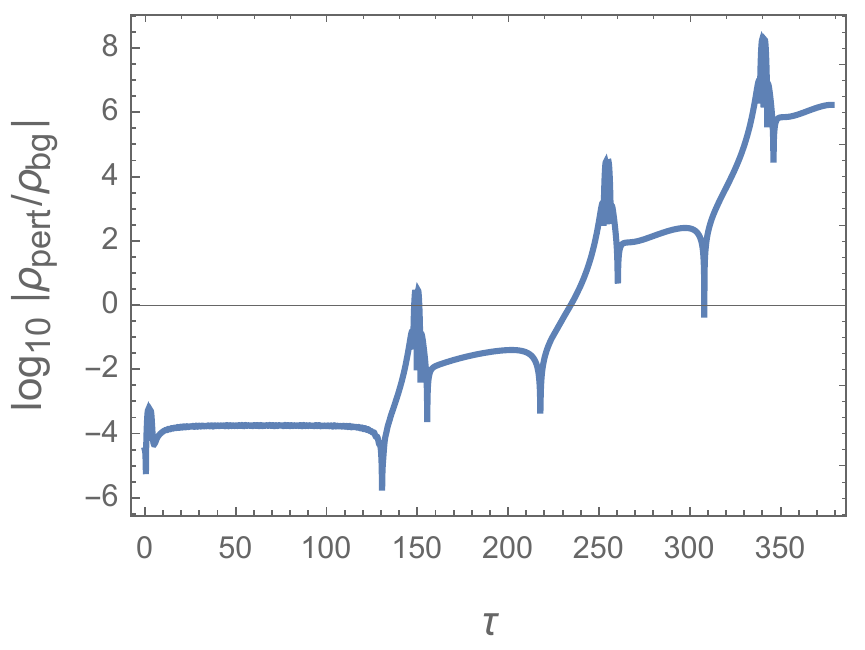}
\end{center}
\caption{(Left) Evolution of a highly--tachyonic fast--growing mode function as a function of time (blue) as compared to the background evolution (orange). The mode function spends most of the time in tachyonic growth, associated with background field values within the potential plateau. (Right) Energy density growth for the same mode.}
\label{fig:mode_exp_growth}
\end{figure}

\noindent The recurrent incursion of the inflaton field in the inflationary plateau has a dramatic effect in the production of Higgs excitations. To see this explicitly let us consider the evolution equation for the Higgs perturbations $q_k$,
\begin{equation} \label{eq:coupled_scalar_eq}
	\ddot{q}_k + 3H\dot{q}_k + \left( \frac{k^2}{a^2} + U''(\chi) \right) q_k 
	= \frac{2}{a^3} \frac{d}{dt} \left( a^3 H \epsilon_H \right) q_k  \, ,
\end{equation}
with $\epsilon_H = \dot{\chi}^2/(2H^2M_P^2)$. The right-hand side of this equation accounts for the coupling to metric perturbations. As confirmed by numerical tests, the $\xi^{-1/2}$ scaling of this mixing term together with the large $\xi$ values needed to reproduce the observed amplitude of the primordial spectrum of density perturbations for reasonable values of $\lambda$  [cf.~Eq.~\eqref{eq:cmb_derived_quantities}] allows to safely approximate Eq.~\eqref{eq:coupled_scalar_eq} as
\begin{gather}
\label{eq:perturbation_eom}
	\ddot{q}_k + 3H\dot{q}_k + \left( \frac{k^2}{a^2} + U''(\chi) \right) q_k = 0 \,. 
\end{gather}
As we did before, we will reduce the dependence on the couplings to a minimum by introducing rescaled variables
\begin{equation} \label{eq:rescaled_variablesQ}
\begin{split}
	Q_\kappa \equiv \, \left( \frac{\lambda}{\xi} \right)^\frac{1}{4} q_k \, , \hspace{10mm}
	\kappa \equiv \sqrt{\frac{\xi}{\lambda}} \frac{k}{M_P} \,, 
	\hspace{10mm}
	E \equiv \frac{\xi}{\sqrt{\lambda} M_P} H \, , \hspace{10mm}
	\tau \equiv \sqrt{\frac{\lambda}{\xi}} M_P \,t\,, 
\end{split}
\end{equation}
such that Eq.~\eqref{eq:perturbation_eom} becomes
\begin{gather}
\label{eq:perturbation_eom_rescaledQ}
	Q''_\kappa + 3\frac{E}{\sqrt{\xi}}Q'_\kappa + \omega_\kappa^2 Q_\kappa = 0 \, , \hspace{20mm} \omega_\kappa^2 \equiv \frac{\kappa^2}{a^2} + u''(y)\,. \end{gather}
\begin{figure}
\begin{center}
\includegraphics[scale=0.8]{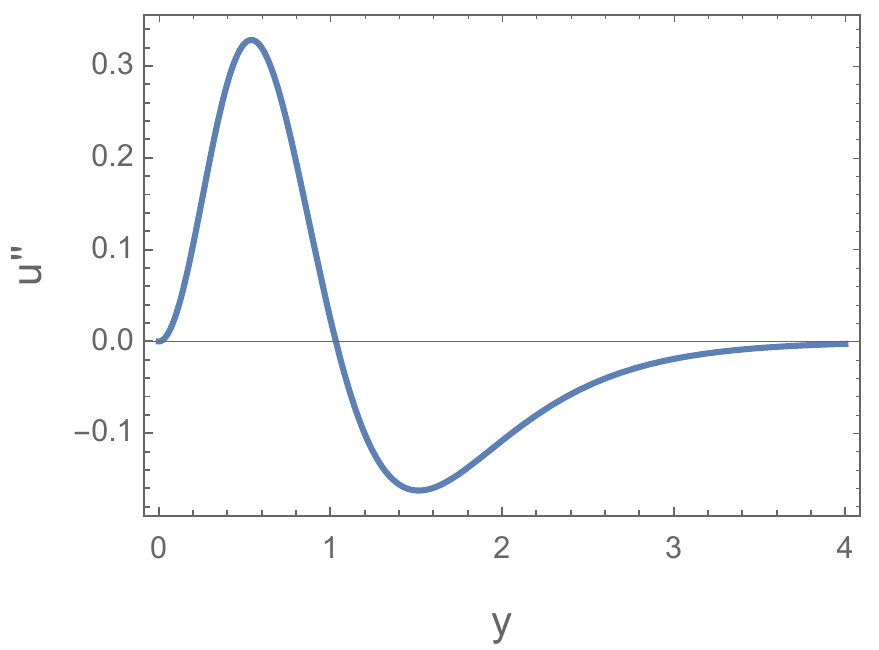}
\includegraphics[scale=0.8]{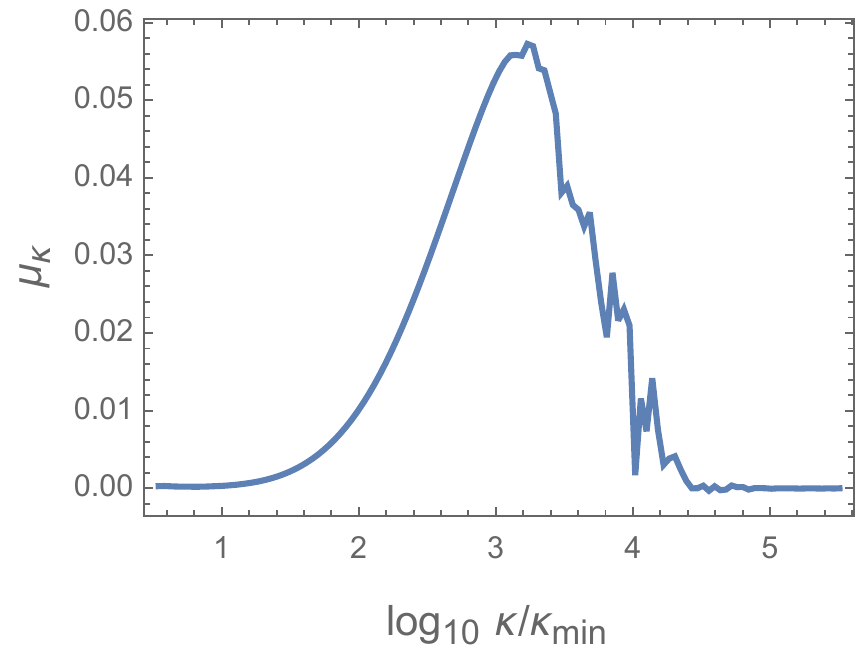}
\end{center}
\caption{(Left) Second derivative of the rescaled Higgs potential $u(y)$ as a function of the dimensionless field variable $y$.
(Right) Growth index $\mu_\kappa$ for different $\kappa$-modes on a logarithmic scale and $\xi=10^8$. The left-hand side of the maximum corresponds to highly tachyonic modes. The oscillations in the right-hand side account for subleading parametric--resonance effects. The parameter $\kappa_\lmin$, corresponding to the wave number that leaves the Hubble radius at the end of inflation, is defined in Eq.~\eqref{eq:kappa_tachyonic}.}
\label{fig:growth_indices}
\end{figure}
The second derivative of the rescaled Higgs potential $u(y)$ as a function of the dimensionless field variable $y$  is shown in the left  hand-side of Fig.~\ref{fig:growth_indices} . As clearly appreciated in this figure, the ``effective mass term" $u''(y)$ in Eq.~\eqref{eq:perturbation_eom_rescaledQ} becomes negative for $y> \frac{1}{2}\ln \left(4+\sqrt{15}\right)\simeq 1$, allowing for \textit{sub-Hubble modes} with momenta 
\begin{equation} \label{eq:kappa_tachyonic}
	\kappa_\lmin < \kappa < \kappa_\lmax \, , \hspace{15mm} \kappa_\lmin \equiv \frac{aE}{\sqrt{\xi}}
	\, , \hspace{15mm}  \kappa_\lmax \equiv a\sqrt{|u''_\lmin|} \approx 0.4 a \,,
\end{equation}
to become tachyonic ($\omega_\kappa^2<0$) during the background field oscillations. The evolution of a typical tachyonic mode following from the numerical solution of Eqs.~\eqref{eq:background_eom_rescaled} and \eqref{eq:rescaled_variablesQ} with vacuum initial conditions
$Q_\kappa =1/(a\sqrt{\kappa})$ and $Q'_\kappa = -i\kappa Q_\kappa/a$ is  shown in the left-hand side of Fig.~\ref{fig:mode_exp_growth}. We observe that the mode is ``locked'' to the background solution such that, when the background field $y$ is on the plateau, the Higgs perturbation $Q_\kappa$ approaches zero, passes it, and then grows exponentially; and when $y$ crosses zero, the derivative of $Q_\kappa$ changes quickly and its growth turns again into decline. However, the net result is an exponential growth
\begin{equation} \label{eq:growth_index}
	Q_\kappa \propto \exp \mu_\kappa \tau \,.
\end{equation}
The momentum dependence of the growth index $\mu_\kappa$ in this expression is shown in the right--hand side of Fig.~\ref{fig:growth_indices}.  The peak in this plot is the result of two competing effects. On the one hand, modes with large $\kappa$ tend to spend more time in the tachyonic region, enhancing the  growth.\footnote{This can be easily understood by examining  the evolution of the zero momentum mode $Q_0$. This mode is just a perturbation of the background solution. The magnitude of the background field $y$ is limited by the energy conservation equation \eqref{eq:background_eom_rescaled}, and so is the magnitude of the full homogeneous part of the field given by the sum of $y$ and the $Q_0$ mode. It follows that the $Q_0$ mode cannot be amplified in preheating but instead oscillates with a constant amplitude like $y$. The situation is, however, not stable: small $\kappa$ modes start to quickly deviate from the $Q_0$ solution and can be amplified exponentially.  Initially $u''<0$ and these solutions grow. The $\kappa \neq 0$ solution grows a little slower because its $\omega_\kappa^2$ is not as negative, but the difference is almost negligible at this stage. When $y$ rolls over small values where $u''$ has positive peaks, the derivative of the mode function changes suddenly and the mode functions start to decrease. Since the $\kappa \neq 0$ mode did not grow quite as strongly earlier, its derivative is plunged by this jump to a lower value than that of $Q_0$. This effect is amplified by the exponential growth of the modes, which start to deviate strongly from now on. While the $Q_0$ mode returns periodically to its initial value, the $\kappa \neq 0$ mode gets amplified more and more in each oscillation. A numerical analysis show that this is an attractor solution, in the sense that the resulting growth index $\mu_\kappa$ is independent of the initial conditions of the mode function, i.e. it only depends on $\kappa$ and the background solution. The bigger $\kappa$ is, the faster the mode deviates from the stable $Q_0$ solution, and the faster it crosses zero after the zero-crossing of $y$ and starts to grow in amplitude again.} On the other hand, large values of $\kappa$ make the frequency $\omega_\kappa^2$ less negative and eventually block the tachyonic instability. 
For the fastest growing mode, the two terms in $\omega_\kappa^2$ are of equal magnitude at the maximum field value $y=y_\lmax$, so $\kappa_\mathrm{fast} \sim a\sqrt{|u''(y_\lmax)|}$, and $\omega^2_{\kappa \, \mathrm{fast}} \sim -\kappa_\mathrm{fast}^2/a^2$. Taking into account that this mode spends most of its time in tachyonic amplification, we can obtain the following estimate for the maximum $\mu_{\kappa}$ value displayed in Fig.~\ref{fig:growth_indices},
\begin{equation} \label{eq:biggest_growth_index}
	\mu_{\kappa \, \mathrm{fast}} \sim |\omega_{\kappa \, \mathrm{fast}}| \sim \frac{\kappa_\mathrm{fast}}{a} \sim \left(\frac{6}{\xi}\right)^{1/4} I^{1/2}_H(\infty)  \, ,
\end{equation}
where in the last step we have made use of \eqref{eq:initial_ymax}.
Combining this expression with the definitions in \eqref{eq:rescaled_variablesQ} and taking into account \eqref{eq:cmb_derived_quantities}, the kinetic energy of this mode can be written as 
\begin{equation} \label{eq:higgs_kinetic_energy}
	\frac{k_\mathrm{fast}}{a} \sim 10^{-5} \xi^{-1/4} M_P  \,.
\end{equation}
For relevant $\xi$ values, this momentum scale is significantly larger than the Higgs vacuum expectation value determining the mass of the Standard Model particles at low energies. Therefore, the created degrees of freedom will be highly relativistic once reheating is over. To estimate when this happens, let us consider the ratio between the energy density of Higgs perturbations $\rho_{\rm pert}$ and the background energy density $\rho_B$,
\begin{equation}\label{eq:higgs_perturbation_energy0}
	\frac{\rho_\mathrm{pert}}{\rho_\mathrm{B}} = \frac{1}{3H^2M_P^2}\int \frac{dk^3}{(2\pi)^3} \frac{1}{2}\left[|\dot{q}_k|^2  +\left( \frac{k^2}{a^2} + U''(\chi) \right) |q_k|^2  \right] \,,
\end{equation}
which in terms of the rescaled variables \eqref{eq:rescaled_variablesQ} can be written as\footnote{We assume here a cutoff regularization for the perturbation energy density. The customary way to regularize the momentum integral is to subtract from it the infinite adiabatic vacuum contribution, or in other words, calculate the instantaneous particle number density from a WKB-approximation, multiply it by the energy of the particle, and sum over the modes. However, the tachyonic modes are not adiabatic at any point in their evolution and the concepts of adiabatic vacuum and particle number are ill-defined. However, because of the exponential amplification of the tachyonic modes, their contribution to the energy density is much bigger than any expected vacuum contribution, so it seems reasonable to neglect any regularization for them. Modes with $\kappa > \kappa_\lmax$ \textit{can} be regularized with an adiabatic subtraction, but their residual contribution to the energy density is much smaller than that of the tachyonic modes, so we simply neglect them by using the cut-off.}
\begin{equation}\label{eq:higgs_perturbation_energy}
	\frac{\rho_\mathrm{pert}}{\rho_\mathrm{B}} =\frac{\lambda}{3E^2}
	\int_{\log  \kappa_\lmin}^{\log \kappa_\lmax} \frac{d \log \kappa \, \kappa^3}{4\pi^2} \left[|Q_\kappa'|^2+\omega_\kappa^2 |Q_\kappa|^2   \right] \, . 
\end{equation}
A simple estimate of the number of oscillations $n_\mathrm{osc}$ needed for this quantity to reach a fraction $X$ can be obtained by considering only the fastest growing mode $\kappa_{\rm fast}$ and the oscillation period $2\Delta\tau$ following from Eq.~\eqref{eq:semiosc_time}.  We get 
\begin{equation} \label{eq:N_oscillations}
X =\frac{\lambda}{3E^2}	 \frac{\Delta \ln \kappa \, \kappa_\mathrm{fast}^3}{2\pi^2} \frac{\kappa_\mathrm{fast}^2}{a^2}  \frac{1}{a^2\kappa_\mathrm{fast}}  \exp\left[2\frac{\kappa_\mathrm{fast}}{a}n_\mathrm{osc}\,2\Delta \tau\right]\,,
\end{equation}
or equivalently\footnote{Remarkably, this expression does not depend on the precise value of the model parameters.}
\begin{equation}
	n_\mathrm{osc} =\frac{1}{4\pi}\log X+\frac{1}{4\pi} \log\left(\frac{\xi}{\lambda}\right)+ \frac{1}{4\pi} \log \left( \frac{\pi^2
\,}{12\,I^2_H(\infty) \Delta \ln \kappa} \right) \approx 1.4\,,
\end{equation}
where in the last step we have made use of Eq.~\eqref{eq:cmb_derived_quantities} with $N_*\approx 50$ and set $\Delta \, \ln \kappa = 1$ and $X=0.1$.\footnote{The choice $X=0.1$ is motivated by several numerical studies on the lattice \cite{Figueroa:2015rqa,Enqvist:2015sua,Repond:2016sol,Figueroa:2016wxr}. Beyond this value, the backreaction of the created particles on the inflaton condensate is typically important and the evolution of the system cannot captured by simple analytical techniques.} Although this result should be understood just as an order-of-magnitude estimate,  it clearly illustrates that the  energy density of created excitations will reach that of the background in a few oscillations, at most. Note also that, even if the additional depletion of the condensate due to particle production is accounted for in the background evolution, the field $y$ will continue to explore the tachyonic region of the potential for several oscillations. Indeed, solving Eq.~\eqref{eq:initial_ymax_condition0} for sizable energy losses $\Delta E_i=0.1 E_i, \, 0.2 E_i, \, 0.3 E_i$ we still get amplitudes $y_{\rm max}=2.2, \, 1.8, \, 1.6$ within the tachyonic regime. This is also confirmed by numerical checks. 
\begin{table}
\renewcommand*{\arraystretch}{1.3}
\begin{center}
\begin{tabular}{c|ccc|ccc}
\hline\hline
\rule{0pt}{4ex}	 & \multicolumn{3}{c|}{Peak} & \multicolumn{3}{c}{Plateau} 	\\ 
\hspace{2mm}  $\xi$\hspace{2mm}	&  \hspace{2mm}  $ n_\mathrm{osc}$ \hspace{2mm}   &\hspace{2mm} $\Delta N$\hspace{2mm}   & \hspace{2mm} $\frac{d \ln \rho_\mathrm{pert}}{dn_\mathrm{osc}}$ \hspace{2mm}   &\hspace{2mm}  $n_\mathrm{osc}$ \hspace{2mm}  & \hspace{2mm}  $\Delta N$ \hspace{2mm} & \hspace{2mm} $ \frac{d \ln \rho_\mathrm{pert}}{dn_\mathrm{osc}}$ \hspace{2mm} \\  \hline
\rule{0pt}{4ex}   \hspace{2mm}  	$10^5$ \hspace{2mm}  &\hspace{2mm}   $1.75$\hspace{2mm}   &\hspace{2mm}   $0.10$ \hspace{2mm}   & \hspace{2mm}  $10.3$\hspace{2mm}   & $2$ &\hspace{2mm}   $0.07$\hspace{2mm}   &\hspace{2mm}   $10.1$ \hspace{2mm}  \\
\hspace{2mm}  $10^6$\hspace{2mm}   &\hspace{2mm}   $1.25$ \hspace{2mm}  & \hspace{2mm}  $0.04$ \hspace{2mm}   &\hspace{2mm}   $12.7$ \hspace{2mm}  & \hspace{2mm}  $1.5$ \hspace{2mm}  & \hspace{2mm}  $0.03$\hspace{2mm}   &\hspace{2mm}   $12.3$ \hspace{2mm}  \\
\hspace{2mm}  $10^7$ \hspace{2mm}  &\hspace{2mm}   $1.25$\hspace{2mm}   & \hspace{2mm}  $0.02$\hspace{2mm}   & \hspace{2mm}  $15.0$ \hspace{2mm}  &\hspace{2mm}   $1.5$\hspace{2mm}   &\hspace{2mm}   $0.02$ \hspace{2mm}  & \hspace{2mm}  $14.2$\hspace{2mm}   \\
\hspace{2mm}  $10^8$\hspace{2mm}   &\hspace{2mm}   $0.75$\hspace{2mm}   & \hspace{2mm}  $0.007$\hspace{2mm}   &\hspace{2mm}   $17.1$ \hspace{2mm}  & \hspace{2mm}  $1.5$\hspace{2mm}   & \hspace{2mm}  $0.009$ \hspace{2mm}  & \hspace{2mm}  $16.0$\hspace{2mm}   \\
\hspace{2mm}  $10^9$\hspace{2mm}   &\hspace{2mm}   $0.75$\hspace{2mm}   & \hspace{2mm}  $0.004$\hspace{2mm}   &\hspace{2mm}   $19.3$ \hspace{2mm}  & \hspace{2mm}  $1$\hspace{2mm}   & \hspace{2mm}  $0.003$ \hspace{2mm}  &\hspace{2mm}   $17.9$ \hspace{2mm}  \\	
 \hline\hline
\end{tabular}
\end{center}
\caption{Results of tachyonic Higgs preheating for different $\xi$ values, with $n_\mathrm{osc}$ the number of background oscillations needed for the energy density in the Higgs excitations to exceed $10\%$ of the background energy density, $\Delta N$ the corresponding number of e-folds and $d\ln \rho_\mathrm{pert}/dn_\mathrm{osc}$ a measure of the perturbation growth rate, fitted to the first 5 semioscillations. These quantities are reported both for plateau and peak energy densities (see main text). The oscillation number $n_\mathrm{osc}$ takes half- and full-integer values for the plateau which is reached after each semioscillation, and $1/4$- and $3/4$-values for the peaks which occur when the background field crosses zero, i.e. in the middle of a semioscillation. We see that preheating through this channel is very fast and efficient.}
\label{tab:higgs_production}
\end{table}

For accurate results, we solved the mode equation \eqref{eq:perturbation_eom_rescaledQ} for different $\xi$ values and a handful of modes placed equidistantly on a logarithmic scale between the momenta $\kappa_\lmin$ and $\kappa_\lmax$ in \eqref{eq:kappa_tachyonic}, starting from adiabatic vacuum conditions during inflation. For each $\xi$ value, we summed over the modes to get the ratio \eqref{eq:higgs_perturbation_energy} and determined the number of oscillations needed for it to reach $10\%$ when evaluated at $y_\lmax$ (this is the typical value in time) and at $y\approx0.54$ (there is a strong peak in the perturbation energy density here). The self-coupling $\lambda$ was fixed case by case by imposing the normalization condition \eqref{eq:cmb_derived_quantities} with $N_*\approx 50$. Note, however, that the linear dependence of the ratio  \eqref{eq:higgs_perturbation_energy} on this coupling makes its impact on the analysis minimal as compared with the exponential growth of fluctuations. The numerical results  in Table \ref{tab:higgs_production} confirm the analytical estimates: tachyonic preheating is very violent in Palatini Higgs inflation and leads to the full depletion of the condensate in just a few oscillations of the background field $y$, with less than one e-fold spent in the preheating stage.

\subsection{Gauge boson production} \label{sec:gauge_boson_production}

 The explosive production of longitudinal gauge bosons due to the 'Riemannian spike' advocated in Refs.~\cite{DeCross:2015uza, DeCross:2016fdz, DeCross:2016cbs, Ema:2016dny, Sfakianakis:2018lzf}
 is absent in the Palatini formulation. Indeed, a simple inspection of Eq.~\eqref{relchih} reveals that nothing special happens at the transition value $h=M_P/\xi$. Both $\dot{\chi}$ and $\dot{h}$ evolve smoothly during the background field oscillations.
 This implies that the simple analysis in Refs.~\cite{GarciaBellido:2008ab,Bezrukov:2008ut,Repond:2016sol} is enough in the Palatini formulation. The friction terms for transverse and longitudinal vector modes are still not equal \cite{Ema:2016dny, Sfakianakis:2018lzf}, but the difference is irrelevant for all practical purposes, as can be easily seem by performing a proper rescaling of the fields leaving behind additional mass terms of order $H^2$, i.e. suppressed for modes well inside the Hubble radius. 
 
 The above reasoning allows us to describe the gauge boson modes as scalar degrees of freedom $B_k$ prior to the onset of backreaction \cite{GarciaBellido:2008ab,Bezrukov:2008ut,Repond:2016sol}. The corresponding evolution equation in momentum space takes the form
\begin{equation}\label{Bmodeeq}
	\ddot{B}_k + 3H\dot{B}_k + \left( \frac{k^2}{a^2} + \frac{g^2}{4\xi^2}\tanh^2 \left( \frac{\sqrt{\xi}\chi}{M_P} \right) \right) B_k  = 0 \, ,
\end{equation}
with  $g=g_2,g_2/\cos\theta_W$ for the $B=W,Z$ bosons, $\theta_W=\tan^{-1}(g_1/g_2)$ the weak mixing angle and $g_1$ and $g_2$ the $U(1)_Y$ and $SU(2)_L$ gauge couplings. In order to reduce the coupling dependence on this equation to a minimum, let us define a rescaled set of variables,
\begin{equation} \label{eq:rescaled_variablesW}
\begin{split}
	A_\kappa \equiv \, \left( \frac{\lambda}{\xi} \right)^\frac{1}{4} B_k \,, \hspace{10mm}
	\kappa \equiv \sqrt{\frac{\xi}{\lambda}} \frac{k}{M_P} \,, 
	\hspace{10mm}
	E \equiv \frac{\xi}{\sqrt{\lambda} M_P} H \, , \hspace{10mm}
	\tau \equiv \sqrt{\frac{\lambda}{\xi}} M_P \,t \,.
\end{split}
\end{equation}
In terms of these quantities, the equation of motion \eqref{Bmodeeq} becomes 
\begin{equation}
\label{eq:gauge_boson_mode_eq_rescaled}
	A''_k + 3\frac{E}{\sqrt{\xi}}A'_k + {\tilde \omega}_\kappa^2 A_k = 0 \, ,
\end{equation}
with 
\begin{equation}
\label{eq:freqA}
{\tilde \omega}_\kappa^2\equiv  \frac{\kappa^2}{a^2} + m_A^2\,, \hspace{20mm} m_A^2 \equiv  \frac{g^2}{4\lambda}\tanh^2 y\, .
\end{equation}
 The friction term in Eq.~\eqref{eq:gauge_boson_mode_eq_rescaled} can be removed by performing a field redefinition $A_\kappa\rightarrow a^{-3/2} A_\kappa$, to obtain\footnote{The redefinition introduces terms proportional to $H^2$ and $\ddot a/a$ that can be safely neglected at scales below the horizon.}
\begin{equation}\label{eq:gauge_boson_mode_eq_rescaled_no_frition}
	A''_\kappa + {\tilde \omega}_\kappa^2 A_\kappa = 0 \, .
\end{equation}
For $g^2 \gg \lambda$, the effective mass $m_A^2$ in \eqref{eq:freqA} significantly exceeds the background oscillation frequency and particle production is limited to a very restricted field interval $|y| \ll y_\lmax$, where the adiabaticity condition $m'_A \ll m_A^2$ is violated \cite{Kofman:1997yn}. This allows us to expand Eq.~\eqref{eq:gauge_boson_mode_eq_rescaled_no_frition} around $y=0$ to get 
\begin{equation} \label{eq:B_mode_eq_scaled_pala}
	-\frac{d^2 A_\kappa}{ds^2} - s^2 A_\kappa = K^2 A_\kappa \, ,
\end{equation}
with 
\begin{equation}
	s \equiv M\tau \,,\hspace{10mm} M \equiv \left(\frac{3}{2\lambda}\right)^{1/4} \sqrt{gE} \, ,\hspace{10mm}  K \equiv \frac{\kappa}{\kappa_{\rm max}}  \, , \hspace{10mm} \kappa_\lmax \equiv a\left(\frac{3 g^2E^2}{2\lambda}\right)^{1/4} \,.
\end{equation}
Equation \eqref{eq:B_mode_eq_scaled_pala} can be formally interpreted as the Schr\"odinger equation for a particle crossing a periodic quadratic potential. Using a WKB approximation, we can compute the number of particles after the $j$-th scattering, $n_{k}(j^+)$, in terms of the number of particles just before that scattering, $n_{k}(j^-)$. After some trivial algebra, we get \cite{Kofman:1997yn}
\begin{equation}\label{eq:nk}
	n_\kappa(j^+) = C(\tau_j) + \left(1+2C(\tau_j) \right)n_\kappa(j^-) +2\cos\theta_{j-1} {\sqrt{C(\tau_j)\left[C(\tau_j)+1\right]}}\sqrt{n^2_\kappa(j^-)+n_\kappa(j^-)} \, ,
\end{equation}
with 
\begin{equation}\label{eq:c_j}
	C(\tau_j) \equiv e^{-\pi K^2(\tau_j)} \, ,
\end{equation}
an infrared window function filtering out modes with $\kappa \gtrsim \kappa_\lmax$, and $\theta_j$ an accumulated phase at each scattering. If the phases $\theta_{j}$ among  scatterings are incoherent, we can reduce \eqref{eq:nk} to a phase-averaged relation 
\begin{equation}\label{eq:nk_2}
\left(\frac{1}{2}+n_\kappa(j^+)\right) \simeq {\cal C}(x_j)\left(\frac{1}{2}+n_\kappa(j^-)\right)\,, 
\end{equation}
with enhancing Bose factor ${\cal C}(\tau_j) \equiv 1+2\text{C}(\tau_j)$.

Once produced, the gauge bosons tend to transfer their energy into the Standard Models fermions ($f$) through decays ($A\rightarrow f\bar f$) and  annihilations ($A A\rightarrow f\bar f $). The decay channels are expected to be the dominant processes at early times, when the gauge boson number densities are small. Let us assume this to be the case for the typical number of oscillations we are interested in. In that case, the gauge boson occupation numbers just before the $j$-th scattering can be written as
\begin{equation}\label{eq:nk_decay}
n_\kappa(j^-) =n_\kappa((j-1)^+)e^{-\langle\Gamma_A\rangle_{j-1} \Delta \tau} \,,
\end{equation} 
with $\left\langle\Gamma_A\right\rangle_j$ the mean decay width of the $W$ and $Z$ bosons  between two consecutive zero-crossings, namely \cite{Cheng:1985bj}
\begin{eqnarray} \label{eq:decay_W_Z}
	\left\langle\Gamma_{W\rightarrow all}\right\rangle_j &=& \frac{3g_2^2\langle m_W\rangle_j}{16\pi}\frac{dt}{d\tau} \equiv \frac{\gamma_W}{\Delta\tau} I_M(y_{\mathrm{max}\,j}) \, , \\
	\left\langle\Gamma_{Z\rightarrow all}\right\rangle_j &=& \frac{2\text{Lips}\left\langle\Gamma_{W\rightarrow all}\right\rangle_j }{3\cos^3\theta_W}
	\equiv \frac{\gamma_Z}{\Delta\tau} I_M(y_{\mathrm{max}\,j}) \,.
\end{eqnarray}
Here $\text{Lips}\equiv\frac{7}{4}-\frac{11}{3}\sin^2\theta_W+\frac{49}{9}\sin^4\theta_W$ a is \textit{Lorentz invariant phase-space} factor, 
\begin{gather}
\label{eq:gamma_factors}
	\gamma_W \equiv\frac{3g_2^{3}}{32\pi\sqrt{6\lambda}E} \, , \hspace{20mm}
	\gamma_Z \equiv \frac{2\text{Lips}}{3\cos^3\theta_W}\gamma_W \, ,
\end{gather}
and
\begin{gather}
\label{eq:IM}
	I_M(z) \equiv 2\int_0^z dx \tanh x \left( 1 - \frac{\tanh^4 x}{\tanh^4 z} \right)^{-1/2} \approx  \frac{\Delta\tau \tanh z}{\sqrt{2}} 
\end{gather}
for $z \gg 1$. 
Combining Eqs.~\eqref{eq:nk_2} and \eqref{eq:nk_decay}, we obtain 
\begin{equation}\label{eq:iter_nk}
\left(\frac{1}{2}+n_\kappa((j+1)^+)\right)={\cal C}(x_j)\left(\frac{1}{2}+n_\kappa(j^+)\,e^{-\gamma I_M(y_{\mathrm{max}\,j})}\right) \, .
\end{equation}
Taking into account that the occupation numbers are invariant under momentum rescalings, 
$n_k = n_{\kappa(k)}$,  the recursive iteration of this master equation allows to obtain the total number density of  gauge bosons at each zero crossing via the momenta integration
\begin{equation}\label{eq:nB}
	n_B(j^+)=\frac{1}{2\pi^2 a^3(\tau_j)}\int_i^\infty dk\,k^2 n_k(j^+)=\frac{1}{2\pi^2 a^3(\tau_j)}\left(\frac{\lambda}{\xi}\right)^{3/2} \int_i^\infty d\kappa\,\kappa^2 n_\kappa(j^+) \, ,
\end{equation}
with the factor $a^3$ accounting for the relation between coordinate densities and physical densities. The total energy density of the non-relativistic gauge bosons is then\footnote{The factor 2 accounts for the $W^+$ and $W^-$ while the factors 3 reflects the fact that each gauge boson can have one of three polarizations.}
\begin{equation}\nonumber
	\rho_B(j) =\rho_W(j) + \rho_Z(j) \, , \hspace{10mm}
	\rho_W(j) = 2\times 3\, n_W(j^+)\langle m_W \rangle_{j}\, , \hspace{10mm}
	\rho_Z(j) =1\times 3\,n_Z(j^+)\langle m_Z \rangle_{j} \, ,
\end{equation}
with
\begin{equation} \label{eq:gauge_mass_expectations}
	\langle m_W \rangle_{j} = \frac{g_2 M_P I_M(y_{\mathrm{max}\,j})}{2\sqrt{6\xi}E\Delta\tau}  \, , \hspace{15mm}\langle m_Z \rangle_{j} = \frac{\langle m_W \rangle_{j}}{\cos \theta_W} 
\end{equation}
the average gauge boson masses.

The energy density lost in decays is transferred to the fermions, increasing their energy density at each semioscillation by 
\begin{equation} \label{eq:delta_rho_fermions}
	\Delta\rho_F{(j)} = \rho_W(j-1)\left( 1 - e^{-\gamma_W I_M(y_{\mathrm{max}\,j})}\right) + \rho_Z(j-1)\left( 1 - e^{-\gamma_Z I_M(y_{\mathrm{max}\,j})}\right)\,.
\end{equation}
Taking into account that the produced fermions are relativistic, their total energy density can be written as
\begin{equation}\label{eq:rho_fermions}
	\rho_F{(j)} = \sum_{i=1}^{j} 
\left(\frac{a_i}{a_j}\right)^{4}\Delta\rho_F{(i)} \, .
\end{equation}
It turns out that as the gauge bosons are produced, practically all of them decay into fermions during the next semioscillation. Indeed, the product $\langle\Gamma_A\rangle_{j} \Delta \tau $ can be well-approximated by
\begin{equation} \label{eq:decay_approximation}
	\langle\Gamma_A\rangle_{j} \Delta \tau \approx \frac{3g_2^3}{64\sqrt{\lambda}}e^{y_\lmax}
\end{equation}
where we have used $E\approx E_i= 1/\sqrt{12}$, $\tanh y_\lmax \approx 1$ and $\Delta \tau$ as given in Eqs.~\eqref{eq:semiosc_time} and \eqref{eq:IT}. For typical values $g_2 \sim 0.1$, $y \gtrsim 1$, and $g^2 \gg \lambda$,  this quantity is larger than one, $\langle\Gamma_A\rangle_{j} \Delta \tau \gtrsim 1$.\footnote{In the Palatini formulation, both the semioscillation time and the average gauge boson masses are bigger than in the metric case, so gauge boson decay is much more efficient.} This implies that the gauge boson density in a particular mode is reduced practically to zero after a semioscillation, pumped afterwards up to $C(\tau_j)$ through Eq.~\eqref{eq:nk} and then reduced to zero again. This results on a steady transfer of energy to the Standard Model fermions. Neglecting the expansion of the Universe during the first oscillations, the ratio between fermion and background energy densities grows in steps of
\begin{equation} \label{eq:fermion_background_energy_ratio_growth}
\begin{split}
	\frac{\Delta\rho_F{(j)}}{\rho_\mathrm{B}(\tau_j)} =
	& \,\frac{1}{3H^2(\tau_j)}\int_i^\infty \frac{dk\,k^2}{2\pi^2a^3}\sum_{A}  C(\tau_j)  \langle m_A \rangle_j  
	\approx 	\,
	10^{-2} g_2^2\left(\frac{g_2^2}{\lambda}\right)^{1/4} \left(1 +  \frac{1}{2 \cos \theta_W} \right) \,,
\end{split}
\end{equation}
which, depending on the ratio $g_2^2/\lambda$,  could be large, leading to a fast onset of radiation domination.
As a consistency check for this result, we performed a numerical analysis. Starting from zero particle numbers at the end of inflation, we determined the evolution of the energy densities of gauge bosons and fermions by i) solving the background evolution numerically, ii) iterating over the background oscillations for a number of sub-Hubble modes up to $\kappa_\lmax$ and iii) solving for particle production according to Eq.~\eqref{eq:nk}. We varied the non-minimal coupling $\xi$ while choosing the Higgs self-coupling $\lambda$ to satisfy the normalization condition \eqref{eq:cmb_derived_quantities} with $N_*\approx 50$. For the gauge couplings, we followed the running of the Standard Model parameters up to the scale of inflation, obtaining $g_2 \sim 0.5$ and $\sin \theta_W \sim 0.4$. The results are presented in Table \ref{tab:fermion_production}.  As expected, practically all the gauge bosons created at a given zero crossing are turned into fermions during the next semioscillation and the fermion energy density grows linearly.  With our parameter values, the growth step \eqref{eq:fermion_background_energy_ratio_growth} is of the order of a few percent per oscillation, meaning that the fermion energy density will exceed a $10\%$ of the background energy density in just a few oscillations. However, the tachyonic Higgs production discussed in Section \ref{sec:higgs_production} is even faster and exponential, making this depletion mechanism subdominant.
\begin{table}
\renewcommand*{\arraystretch}{1.3}
\begin{center}
\begin{tabular}{c|c|c}
 \hline\hline
\rule{0pt}{4ex}   	\hspace{5mm} $\xi$ \hspace{5mm}  & \hspace{5mm}  $\frac{g_2^2}{\lambda}$ \hspace{5mm}  & \hspace{5mm}  $\langle \frac{\Delta\rho_F}{\rho_\mathrm{B}} \rangle$ \hspace{5mm}  \\ \hline
\rule{0pt}{4ex}    \hspace{5mm} 	$10^5$\hspace{5mm}  & \hspace{5mm} $27789$ \hspace{5mm} &\hspace{5mm}  $0.064$ \hspace{5mm} \\
\hspace{5mm} 	$10^6$\hspace{5mm}  &\hspace{5mm}  $2814$ \hspace{5mm} &\hspace{5mm}  $0.037$ \hspace{5mm}  \\
\hspace{5mm} 	$10^7$\hspace{5mm}  &\hspace{5mm}  $285$ \hspace{5mm} &\hspace{5mm}  $0.022$ \hspace{5mm} \\
\hspace{5mm} 	$10^8$\hspace{5mm}  &\hspace{5mm}  $29$\hspace{5mm}  &\hspace{5mm}  $0.014$ \hspace{5mm} \\
\hspace{5mm} 	$10^9$ \hspace{5mm} &\hspace{5mm}  $2.9$\hspace{5mm} &\hspace{5mm}  $0.014$ \hspace{5mm} \\	
\hline\hline 
\end{tabular}
\end{center}
\caption{Results of gauge boson and fermion production for different $\xi$-values. Ratio $g_2^2/\lambda$ is large for all $\xi < 10^9$, but the $\xi=10^9$ result is on edge of validity of our analysis. The third column shows how much the fermion energy density grows per semioscillation, averaged over the first 30 semioscillations of preheating.}
\label{tab:fermion_production}
\end{table}
\section{Impact on cosmological observables} \label{sec:impact}
The analysis presented in Section \ref{sec:particle_production} reveals that the dominant preheating mechanism in Palatini Higgs inflation is the tachyonic production of Higgs excitations. As summarized in Table \ref{tab:higgs_production}, this process turns the background field energy density into Higgs quanta in a negligible number of e-folds. In practice this means that preheating in Palatini Higgs inflation is essentially instantaneous and almost all of the background energy density at the end of inflation is turned into radiation. The reheating temperature
\begin{equation} \label{eq:T_reh}
	T_\mathrm{RH} = \left(\frac{30\,\lambda}{4\pi^2\xi^2g_{*\, \mathrm{RH}}}\right)^{1/4} M_P\,,
\end{equation}
is then close to the energy scale of inflation, with $g_{*\, \mathrm{RH}}=106.75$ the effective number of relativistic degrees of freedom for the Standard Model particle content and the caveat that we have not considered the details of the thermalization process.\footnote{The energy density leading to a thermal description is expected to be slightly lower than $T_{\rm RH}$ .} 

Armed with the results of the previous sections, we can calculate the inflationary observables to a higher order of accuracy than before. These observables depend on the number of e-folds of inflation needed to solve the flatness and horizon problems given a particular cosmological history. Following the standard procedure, we require 
\begin{equation} \label{eq:efold_formula}
	1=a_0=\frac{a_0}{a_\mathrm{RH}}\frac{a_\mathrm{RH}}{a_\mathrm{end}}\frac{a_\mathrm{end}}{a_*}a_*
	=	\left(\frac{g_{*s\,\mathrm{RH}}}{g_{*s\,\mathrm{now}}}\right)^{1/3} \frac{T_\mathrm{RH}}{T_0}
	\frac{k_*}{H_*}\exp\left(\Delta N+N_*\right) \,,
\end{equation}
with the different subindices referring to the value of the corresponding quantity at the present time (``0"), the end of the reheating stage (``RH"), the end of inflation (``end'')  and the time at which the pivot scale $k_* = 0.05$ Mpc${}^{-1}$ crosses the horizon (``*").  The quantity $\Delta N$ denotes the number of e-folds of reheating, $N_*$ is the number of e-folds of inflation left at the pivot scale, $g_{*s}$ is the effective number of entropy degrees of freedom with $g_{*s\,\mathrm{RH}}=g_{*\, \mathrm{RH}}$ and $g_{*s\,\mathrm{now}}=3.94$ \cite{Husdal:2016haj} and $T_0\simeq 2.7$ K. 

Solving \eqref{eq:efold_formula} for $N_*$ and using Eqs.~\eqref{eq:cmb_derived_quantities}, \eqref{eq:background_eom} and \eqref{eq:T_reh}, together with the condition $\Delta N \ll 1$, we get a relation 
\begin{equation} \label{eq:efold_formula_palatini}
	N_* = 54.9 - \frac{1}{4}\log \xi \, ,
\end{equation}
which is accurate to an integer order in $N_*$. This accuracy  translates into a $10^{-3}$ accuracy on the spectral tilt that exceeds the ${\cal O}(1/N_*)$ accuracy of $n_s$ in \eqref{eq:cmb_observables}. The consistency of the procedure requires therefore to account for ${\cal O}(1/N_*^2)$ corrections by considering second order corrections in slow-roll. In terms of the (potential) slow-roll parameters $\epsilon_V$, $\eta_V$ and $\xi_V$, we have \cite{Stewart:1993bc, Liddle:1994dx}
\begin{equation}
\label{eq:ns_second_order}
	n_s \approx 1 - 6\epsilon_V + 2\eta_V +\frac{1}{3}(44-18c)\epsilon_V^2 + (4c-14)\epsilon_V\eta_V + \frac{2}{3}\eta_V^2 + \frac{1}{6}(13-3c)\zeta_V \, ,
\end{equation}
with $c=4(\gamma + \log2) - 5$ and $\gamma=0.5772\dots$ the Euler-Mascheroni constant. 
\begin{figure}
\begin{center}
\includegraphics[scale=1]{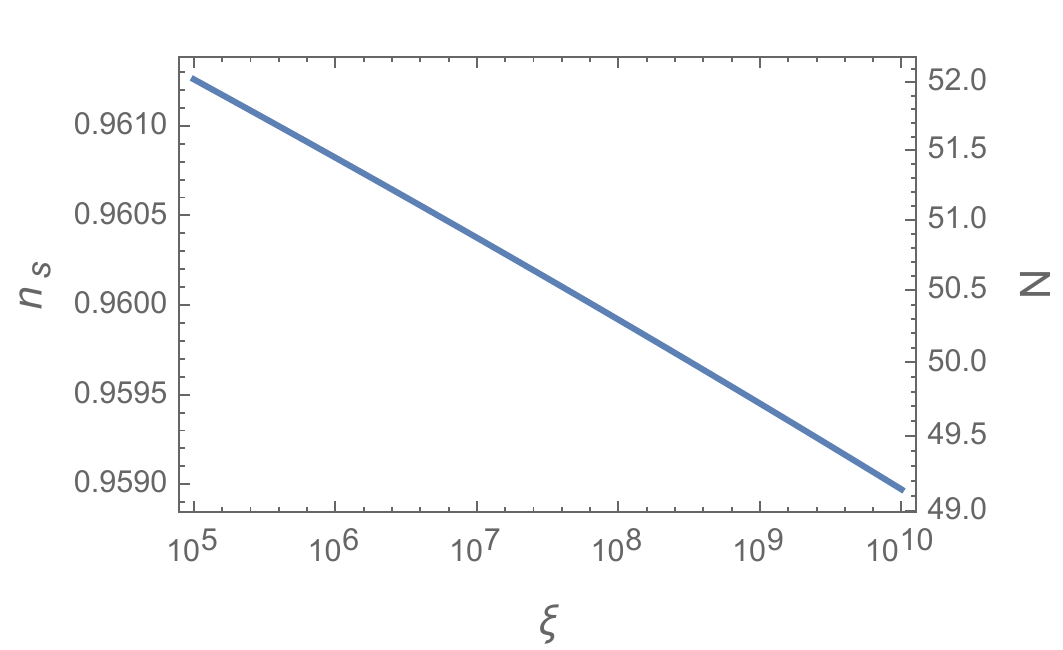}
\end{center}
\caption{Spectral index $n_s$ and number of e-folds of inflation left $N$ at the CMB pivot scale, as functions of $\xi$.}
\label{fig:xi_ns_N}
\end{figure}
Up to ${\cal O} (1/\xi)$ corrections, the slow-roll parameters in Palatini Higgs inflation take the form \cite{Bauer:2008zj}
\begin{equation} \label{eq:palatini_sr_parameters}
	\epsilon_V \approx \frac{1}{8\xi N_V^2} \, ,\hspace{20mm}  \quad \eta_V \approx -\frac{1}{N_V} \, , \hspace{20mm} \quad \xi_V \approx \frac{1}{N_V^2} \, ,
\end{equation}
with 
\begin{equation} \label{eq:sr_efolds}
	N_V \equiv \int_i^\chi \frac{d\chi}{\sqrt{2\epsilon_V}}
\end{equation}
the number of e-folds and $\chi$ the corresponding background field value. However, Eq.~\eqref{eq:sr_efolds} is just an approximation that breaks down near the end of inflation; numerical simulations show that at the $N_* \approx 50$ ballpark, we must correct this number by $N_* \approx N_V + 1.8$, which is accurate to integer order for large $\xi$ values. Combining the above expressions and retaining only the terms up ${\cal O}(1/N_*^2)$, we get
\begin{equation} \label{eq:ns_accurate}
	n_s \approx 1 - \frac{2}{N_V} + \frac{2.8}{N_V^2} \approx 1 - \frac{2}{N_*} - \frac{0.8}{N_*^2} \, ,
\end{equation}
which can be converted to a relation between $n_s$ and $\xi$ through Eq.~\eqref{eq:efold_formula_palatini}, see also Fig.~\ref{fig:xi_ns_N}.
For $\xi=10^6 \dots 10^9$ we get $N_*=50 \dots 51$ and $n_s=0.959 \dots 0.961$. Although still within two--sigma confidence limits, these spectral tilt values are slightly disfavoured by the Planck result $n_s =  0.9653 \pm 0.0041$ \cite{Akrami:2018odb}. Note, however, that, given the expected ${\cal O}(10^{-3})$ accuracy of forthcoming experiments \cite{Martin:2014rqa}, the Palatini predictions could be potentially distinguished from the metric ones in a near future.

In deriving the above results, we made several assumptions to reduce the physical scenario to a baseline model that could be treated with simple analytical and numerical techniques:
\begin{enumerate}
    \item We neglected the potential decay of Higgs excitations into other Standard Model particles. Although the impact of this effect seems difficult to estimate in the absence of a particle-like interpretation for tachyonic modes, we expect it to play a minor role in the analysis, since, contrary to parametric resonance, the tachyonic production does not depend on the number of previously--existing excitations.  
    \item We assumed an instantaneous shift to radiation domination when the Higgs excitations start to dominate, which, although being a reasonable premise, is certainly beyond our simple analysis. Note, however, that this hypothesis seems to be in good qualitative agreement with the lattice simulations performed in Ref.~\cite{Lozanov:2019ylm},  where, for similar potentials, the Universe approaches a radiation dominated regime with equation--of--state parameter $w\simeq 1/3$ almost immediately after the end of inflation.
    \item We restricted ourselves to a tree-level analysis. The qualitative results presented in this paper are expected to hold, however, upon the inclusion of radiative corrections, provided that i) these respect the asymptotic symmetries of the theory at large field values and that ii) the Higgs self-coupling is larger than a critical value at the inflationary scale \cite{Bezrukov:2014ipa,Rubio:2015zia,Fumagalli:2016lls,Enckell:2016xse,Bezrukov:2017dyv,Rasanen:2017ivk,Enckell:2018kkc}.
\end{enumerate}

\section{Conclusions}\label{sec:conclusions}

Higgs inflation is probably the simplest inflationary model, not only because it does not require the introduction of additional scalar fields beyond the one already present in the Standard Model, but also because the strength of the interactions among the inflaton field and the matter sector are experimentally known.

In spite of its simplicity, Higgs inflation is not free 
of caveats. On top of the ambiguities associated with the precise ultraviolet completion of the model, there are additional uncertainties entering the definition of the non-minimal coupling to gravity. The customary choice is a metric formulation of gravity in which the Ricci scalar is computed out of a Levi-Civita connection depending on the metric tensor. Note, however, that an alternative Palatini formulation based on a connection completely unrelated to the metric tensor should not be a priori excluded, since it leads essentially to the same physics than the metric formulation in today's Universe.

The choice of gravitational degrees of freedom can have a strong impact on dynamics. In this paper we performed a detailed study of the preheating stage following the end of Palatini Higgs inflation. We showed that, contrary to the metric case, the depletion of the Higgs condensate is not dominated by the parametric production of Standard Model gauge bosons. The slow decay of the Higgs oscillations after the end of inflation allows the field to periodically return to the plateau of the potential. In this large--field regime, the effective Higgs mass becomes negative, allowing for the exponential creation of Higgs excitations. This entropy production process turns out to be very efficient and leads to the complete depletion of the inflaton condensate in less than one e-fold of expansion. The preheating stage in  Palatini  Higgs  inflation is therefore essentially instantaneous. As compared to the metric case, this reduces the number of e-folds of inflation needed to solve the usual hot big bang problems while leading to a slightly smaller value for the spectral tilt of primordial density perturbations. 

\section*{Acknowledgments}
We thanks Kari Enqvist and Syksy R{\"a}s{\"a}nen for useful discussions. ET is supported by the Vilho, Yrj{\"o} and Kalle V{\"a}is{\"a}l{\"a} Foundation of the Finnish Academy of Science and Letters.

\bibliographystyle{apsrev}
\bibliography{mybiblio}
\end{document}